\documentclass{iaus}
\usepackage{graphicx}

\title[The Ultraluminous X-ray Source in Holmberg IX and its Environment] %% give here short title %%
{The Ultraluminous X-ray Source in \mbox{Holmberg IX} and its Environment}

\author[F. Gris\'e et al.]   %% give here short author list %%
{Fabien Gris\'e$^1$%
  , Manfred W. Pakull$^1$ \break \and Christian Motch$^1$}

\affiliation{$^1$Observatoire Astronomique de Strasbourg,
11 rue de l'Universit\'e, FRANCE \break email: grise@astro.u-strasbg.fr\\[\affilskip]
}

\pubyear{2005}
\volume{230}  %% insert here IAU Symposium No.
\pagerange{1--2}
\date{?? and in revised form ??}
\jname{Populations of High Energy Sources in Galaxies}
\editors{E. J. A. Meurs \& G. Fabbiano, eds.}
\begin{document}

\maketitle

\begin{abstract}
We present optical observations of an ultraluminous X-ray source (ULX) in Holmberg IX, a dwarf galaxy near M81. The ULX has an average X-ray 
luminosity of some $10^{40}$ erg/s. It is located in a huge (400pc x 300pc) ionized nebula being much larger than normal supernova remnants. From the observed emission lines (widths and ratios) we find that the structure is due to
collisional excitation by shocks, rather than by photoionization.

We identify the optical counterpart to be a 22.8 mag blue star ($M_V$=-5.0) belonging to a small stellar cluster. From isochrone fitting of our multi-colour photometry we determine a cluster age of 20 to 50 Myr. We also discovered strong stellar HeII$\lambda$4686 emission (equivalent width of \mbox{10 \AA}) which proves the identification with the X-ray source, and which suggests the presence of an X-ray heated accretion disc around the putative black hole.
\end{abstract}

\firstsection % if your document starts with a section,
\section{Introduction}

%Ultraluminous X-ray sources (ULXs) are a class of non-nuclear X-ray point sources in nearby galaxies having X-ray luminosities of $10^{(39-41)}$ erg/s, i.e. greater than the Eddington luminosity for stellar mass black holes. Three main hypotheses concerning the nature of these objects have been discussed in the literature: genuine stellar mass black holes emitting at super-Eddington rates (Begelman 2002), the much advertised intermediate mass black holes (IMBHs) having $10^2$ to $10^5$ solar masses (Colbert \& Mushotzky 1999), or non-isotropic emission beamed into our line-of-sight (King et al. 2001).\\
The two main hypotheses to explain the high luminosity of ULXs are intermediate mass black holes (IMBHs) having $10^2$ to $10^5$ solar masses (Colbert \& Mushotzky 1999) or non-isotropic emission beamed into our line-of-sight (King et al. 2001).\\
Here, we are interested in one of these object, Holmberg IX X-1, located at a distance of 3.6 Mpc in a dwarf galaxy companion of M81. Miller (1995) discovered the nebula around the position of the X-ray source and proposed that this object was an extremely luminous supernova remnant, but the presence of X-ray variability (La Parola et al. 2001) has shown that it is a compact X-ray source.\\
Our optical observations were carried out in 2003 and 2004 with the 8.2 meter SUBARU telescope on Mauna Kea, Hawaii.

\section{Results and discussion}
\begin{figure}[!h]
  \begin{center}
    \begin{tabular}{cc}
      \resizebox{!}{6cm}{\includegraphics{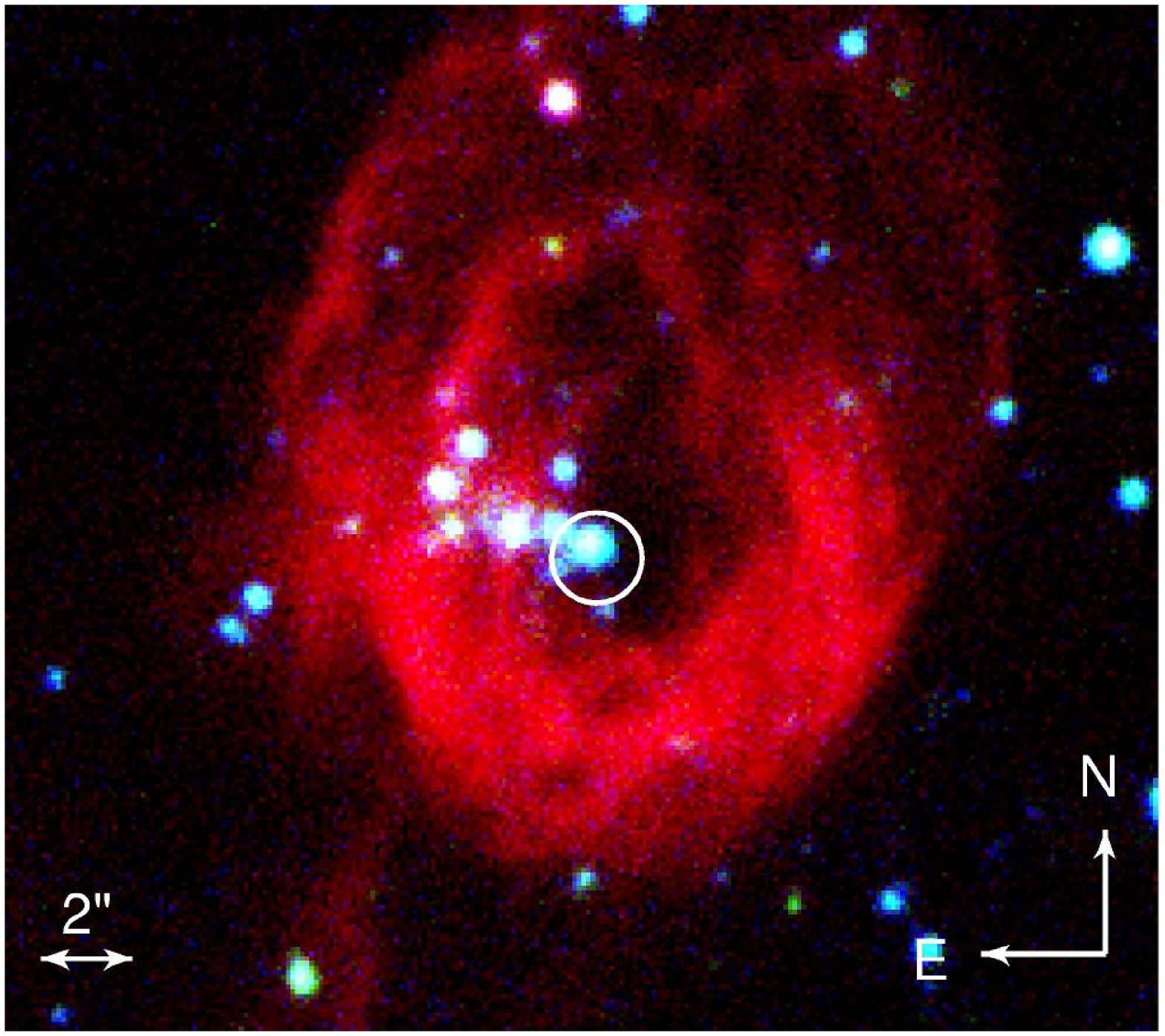}}&
      \resizebox{!}{6cm}{\includegraphics{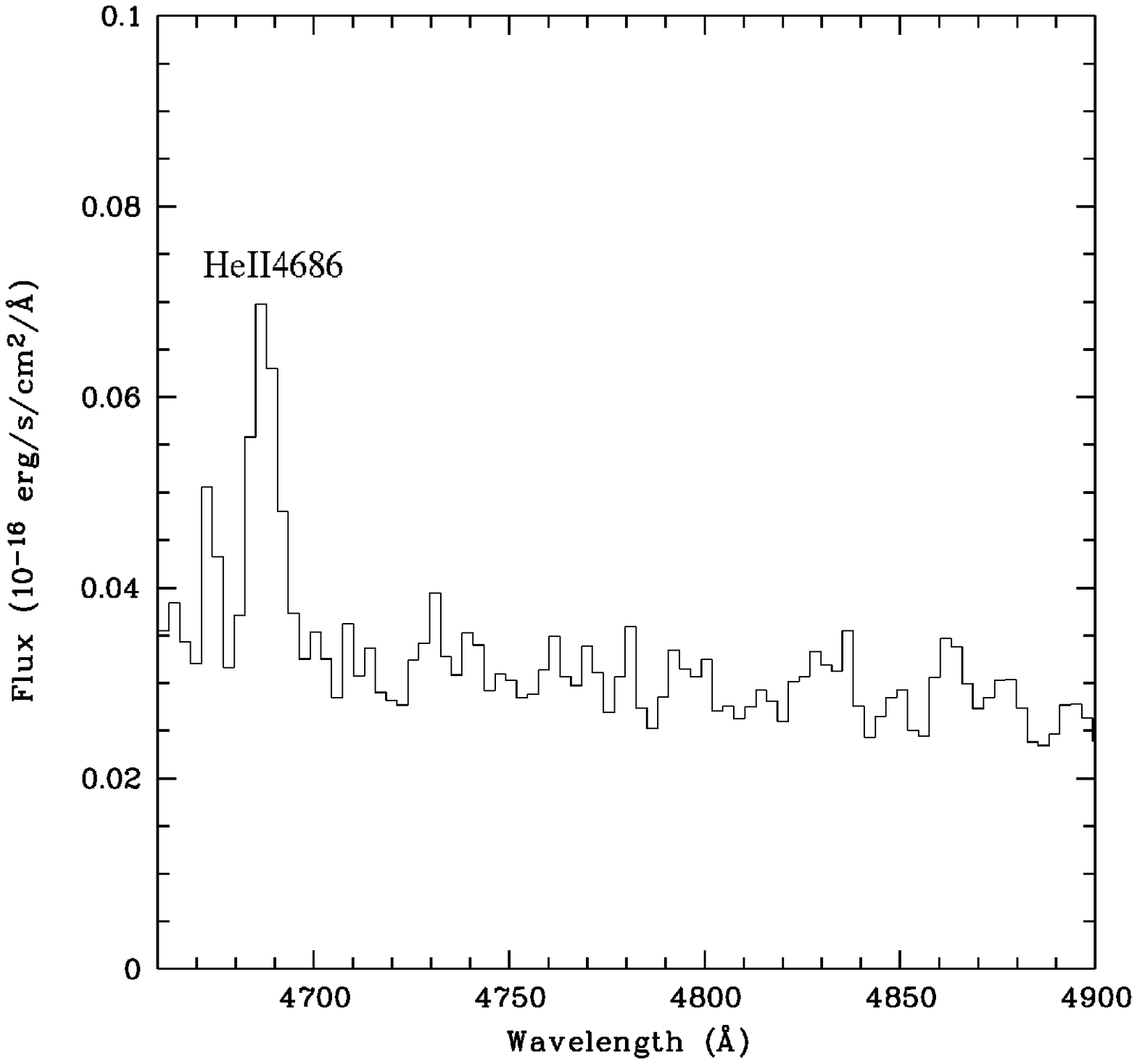}} \\
    \end{tabular}
%\caption[text1]{text2 \\\hspace{\linewidth} text3}
    \caption[]{\\\hspace{\linewidth}
Left : H$_{\alpha}$ (red), V (green) and B-band (blue) multicolor image of the stellar cluster in the local environment of the ULX. The optical counterpart is the bright star in the Chandra error circle.\\
Right : 1D spectrum of the stellar counterpart of Holmberg IX X-1 displaying the HeII$\lambda$4686 emission line.}
    \label{figures}
  \end{center}
\end{figure}

%hoix_heii_1d.jpg

In the huge supernova remnant-like complex, our images reveal that a "blue non-stellar object" seen in previous studies is in fact a star cluster where the brightest members are resolved (Figure \ref{figures}).The most luminous object is a 22.8 mag object, located in the 1" radius Chandra error circle ; we note that in archive HST images this object has a fainter (by 1.9 mag) companion to the west. 

\bigskip

One very interesting result is the discovery of the HeII$\lambda$4686 emission in the brightest star (i.e. not in the fainter companion) with an equivalent width (EW) of some \mbox{10 \AA} (Figure \ref{figures}), proving that it is the optical counterpart of the X-ray source. This is a common feature in luminous massive X-ray binaries, but here with an EW ten to twenty times higher.
This strongly suggests the presence of an accretion disk that is heated by the very luminous X-ray source. This constitutes further evidence against beaming in ULX and opens the possibility to measure the binary orbit from radial velocity observations.\\

Isochrone fitting to our multi-colour photometry has permitted to constrain the age of the star cluster to which the ULX belongs to about 20 to 50 Myr, and to put an upper limit on the total cluster mass - some 10$^3$ M$_{\odot}$. This directly implies that the donor component in the ULX cannot be more massive than 8 M$_{\odot}$, because more massive cluster stars have already exploded as supernovae.\\
%Besides, considering that the object responsible for the formation of the nebula is linked to the stellar cluster, it appears that the progenitor of the supernova cannot be more massive than 10 M$_\odot$, excluding the presence of an IMBH and even of a black hole.\\
%So how can we explain the high initial energy calculated, some 10$^{53}$ erg/s together with the age and large diameter (about 300 pc) of the nebula ? We must invoke the presence of jets, as present in the system SS433 but there is no visible signature at the moment.

%This is in contradiction with the kinematic age of the bubble - some 1 Myr - and opens the road to other hypotheses than IMBHs.

\section{Perspectives}
Optical observations of radial velocity variations of the HeII$\lambda$4686 line in ULXs will hopefully allow to determine the masses of the components in these systems.
This would be a decisive test on whether IMBH are present in ULXs or not. Optical observations of other ULX (in particular NGC 1313 X-2) show strikingly similar characteristics (see contribution at this symposium by Pakull et al.).\\
Future optical observations will be crucial to reveal nature and evolution of the exciting class of ultraluminous X-ray emitters.

\end{document}